\documentclass[12pt]{iopart}

\usepackage{graphicx}
\usepackage{amssymb}
\usepackage{epsfig}

\def\Journal#1#2#3#4{{#1} {#2} (#3) #4}

\def\NPA{Nucl. Phys. A}
\def\PLB{Phys. Lett.  B}

\def\PRC{Phys. Rev. C}
\def\PRD{Phys. Rev. D}

\def\be{\begin{equation}}
\def\ee{\end{equation}}

\newcommand{\ud}{\mathrm{d}}

\begin{document}

\title{Quarkonium production in ultra-relativistic nuclear collisions:
  suppression vs. enhancement}

\author{P.~Braun-Munzinger}

\address{Gesellschaft f\"ur Schwerionenforschung, GSI, 
D-64291 Darmstadt, Germany, and Technical University Darmstadt, D-64289 Darmstadt, Germany}

\begin{abstract}
  After a brief review of the various scenarios for quarkonium production in
  ultra-relativistic nucleus-nucleus collisions we focus on the ingredients
  and assumptions underlying the statistical hadronization model. We then
  confront model predictions for J/$\psi$ phase space distributions with the
  most recent data from the RHIC accelerator. Analysis of the rapidity
  dependence of the J/$\psi$ nuclear modification factor  yields first
  evidence for the production of J/$\psi$ mesons at the phase boundary.
  We conclude with  predictions for charmonium production at the LHC.
\end{abstract}

\vspace{2mm}

\section{General considerations}

Charmonium production is, since the original proposal about its possible
suppression in a Quark-Gluon Plasma (QGP) \cite{satz}, considered as an
important tool to diagnose the fireball produced in ultra-relativistic
nucleus-nucleus collisions.  The original idea of $J/\psi$ 'melting' via Debye
screening \cite{satz} implies rapid production of charmonia in initial hard
collisions and their subsequent destruction in the QGP. For efficient melting
all J/$\psi$ mesons have to be formed well before the QGP temperature has
fallen below T$_D$, the temperature above which screening takes place. A
detailed discussion of this scenario and of the various time scales involved
can be found in \cite{satz1,blaizot}. 

Recent studies of charmonium survival in a hot plasma performed within the
framework of lattice QCD indicate that T$_D$ may be significantly higher than
the critical temperature T$_c$, reaching 2 T$_c$ for J/$\psi$ mesons while
excited states would melt much earlier. If substantiated in unquenched
calculations this could have significant influence on the production pattern
expected in nucleus-nucleus collisions. Here we remark that in such
considerations the widths of charmonia in the QGP should also be taken into
account. A simple estimate of collisional broadening will illustrate this. The
mean free path of the J/$\psi$ in the QGP is the $\lambda =1/(n_p \sigma$.
Since we are interested in temperatures substantially higher than T$_c$ we
assume a QGP with 3 massless flavors plus gluons, leading to a parton density
n$_p$ = 4.25 T$^3$. For the specific case of J/$\psi$ mesons we assume a
J/$\psi$ - parton cross section $\sigma \approx$ 2 mb.  Estimating the
relative velocity v$_{rel}$ of J/$\psi$ vs partons from its thermal velocity
we obtain the in-medium width $\Gamma = v_{rel}/\lambda$. Numerical values
reach $\Gamma$(T=300 MeV) = 320 MeV and $\Gamma$(T=400 MeV) = 760 MeV.  These
large widths imply that most of the charmonia will decay inside the QGP and
thus are not likely reconstructed in an actual experiment. Consideration of
such widths is obviously important if one looks for a characteristic pattern
due to sequential melting of various charmonium states.

Another issue to be considered is 'cold nuclear matter' suppression. Here the
idea is that the 'instantaneously' produced charmonia are partially destroyed
by the passing of the two Lorentz contracted nuclei. While the time scales
involved are such that this may happen at SPS energy (here production and
passing-by times are of the order of 1 fm, \cite{blaizot}), at LHC energy the
passing-by time is about 1/200 fm, and cold nuclear matter effects should be
very small.

\section{Ingredients and assumptions of the statistical hadronization model}

The statistical hadronization model  \cite{pbm1,aa1,aa2} assumes that all
charm quarks are produced in initial hard collisions while charmonia which are
produced early are completely destroyed in the QGP. Cold nuclear matter
effects or destruction by comoving hadrons are consequently not considered.
The entire production of charmonia rather takes place at chemical 
freeze-out, i.e at T$_c$. In this sense charmonia and also all hadrons with
charm are produced by a mechanism similar to that for hadrons containing u, d,
and s quarks, although charm quarks are very far out of chemical equilibrium.
A proposal for a detailed mechanism of hadron production at the phase boundary
can be found in \cite{wetterich}. We  note that a two-component model
including (partial) screening, nuclear absorption, and generation at the phase
boundary was developed in \cite{rapp}.

Under these conditions charmonium production yields scale quadratically with
the number of produced charm quarks, implying little suppression or even
enhancement at the highest energies even though there is complete suppression
in the QGP. We note that, in general, charmonium generation can only take place
effectively if the charm quarks reach thermal (not chemical) equilibrium and
are free to travel over a large distances corresponding to about 1 unit in
rapidity\footnote{For a translation of rapidity into longitudinal distance see
  \cite{ceres_prl}.}, implying deconfinement.

\begin{figure}[ht]
\begin{tabular}{lr}
\begin{minipage}{.48\textwidth}
\vspace{-1cm}
  \centering\includegraphics[width=1.2\textwidth]{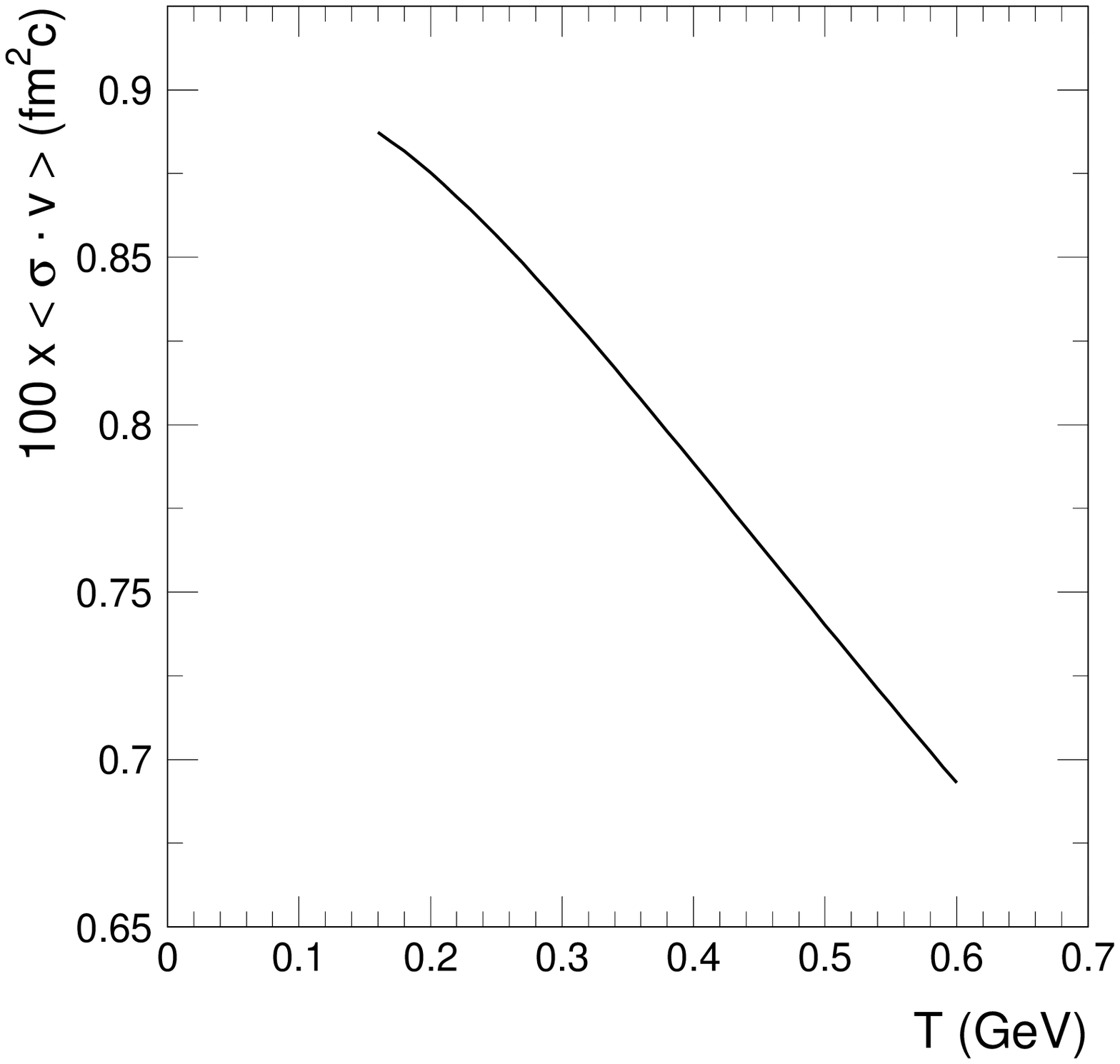}
  \caption{Temperature dependence of the thermal average of the charm
    annihilation cross section as defined in
    the text.}
  \label{aa_fx1}
\end{minipage}  & \begin{minipage}{.48\textwidth}
\vspace{-1cm}
  \centering\includegraphics[width=1.2\textwidth]{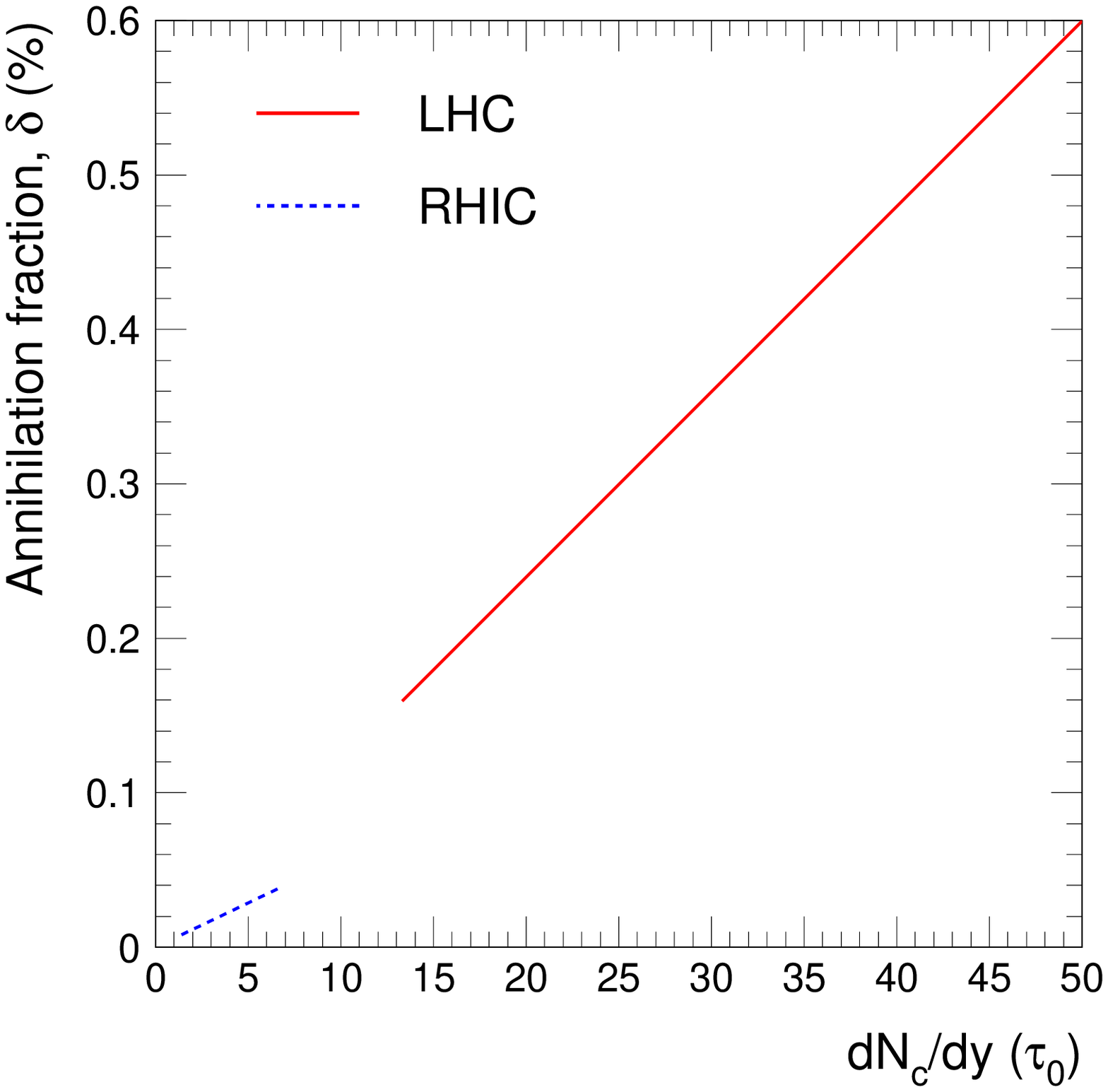}
\caption{Annihilation rate as a function of initial charm rapidity density.}
  \label{aa_fx2}
\end{minipage}
\end{tabular}
\end{figure}

A crucial assumption in the statistical hadronization model is that the number
of charm quarks stays constant 
during the evolution of the plasma. This has been analyzed in some detail in
\cite{aa2}  by evaluation of the rate equation
\be
\frac{dr_{c\bar{c}}} {d\tau} = n_c n_{\bar{c}} \langle
\sigma_{c\bar{c}\rightarrow gg} v_r \rangle,
\ee
where $\langle \sigma_{c\bar{c}\rightarrow gg} v_r \rangle$ is the thermal
average of the annihilation cross section times the relative velocity $v_r$ in
the QGP, and $n_c$ = $n_{\bar{c}}$ is the charm quark density. The quantity
$\frac{dr_{c\bar{c}}} {d\tau}$ is the annihilation rate per volume or the rate
of change of the charm quark density. The total annihilation rate is then
obtained by folding with the temperature evolution of the QGP. Results are
given in Fig.~\ref{aa_fx1}.

For RHIC and LHC energies these estimates imply that charm quark annihilation
in the plasma can be safely neglected. Along the same line, production of
charmonia via uncorrelated charm quark annihilation in the QGP is expected to
fall significantly below the above computed annihilation yield into gluons,
lending strong support to the above interpretation that all quarkonia are
produced late, when the system reaches the critical temperature and
hadronizes. These results are not likely to be changed if annihilation into 3
or more gluons were taken into account. We first note that annihilation into n
gluons are suppressed by a factor $\alpha_s(m_{J/\psi})^{3+n}$, and one can
get an impression of the suppression by comparison of the width of J$/\psi$
(decaying into 3 gluons) with that of the $\eta_c$ (decaying into 2 gluons).
Furthermore, gluons in the QGP acquire thermal masses $\propto$ gT, implying a
further reduction. 

The above discussion underlines the differences between the statistical
approach, 
where, except for corona effects \cite{aa2},  all charmonia are formed
non-perturbatively at T$_c$,  and the kinetic 
model of \cite{the1,the3,yan}, where charmonia are recombined during plasma 
evolution from, in general, uncorrelated charm quarks.

\section{Confrontation of statistical hadronization model  predictions with
  data} 
In the following we base our quantitative comparisons of model predictions to
data on the approach developed in \cite{aa2}. For the production of charm
quarks via initial hard collisions we use the calculations of \cite{cac,cac1},
for RHIC energy and the predictions of \cite{rv1} for LHC energy. All
calculations are performed in the framework of perturbative QCD for
nucleon-nucleon collisions and scaled to nucleus-nucleus collisions with
the appropriate (geometric) number of binary collisions.

\begin{figure}[ht]
  \centering\includegraphics[width=.95\textwidth]{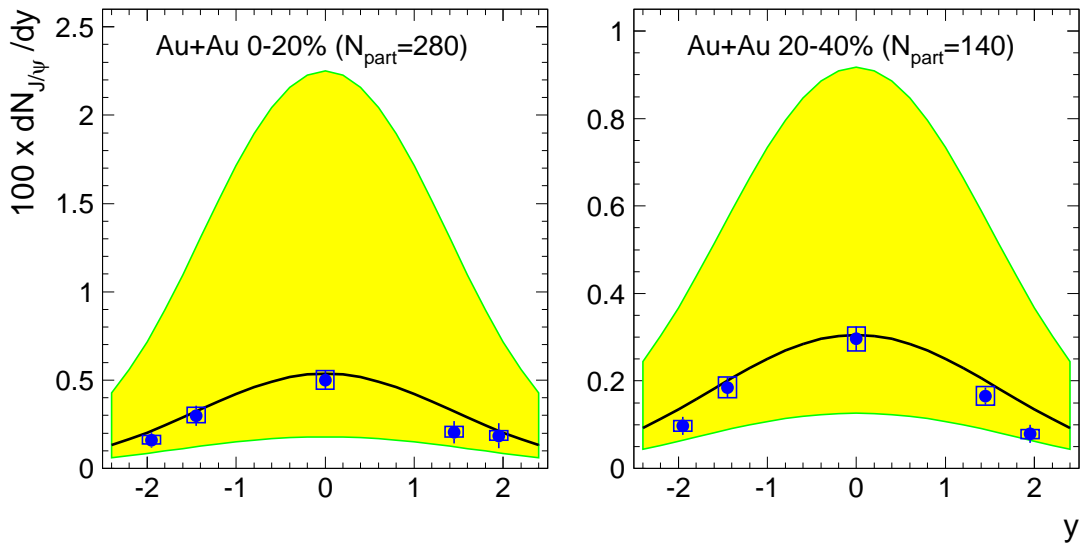}
  \caption{Rapidity dependence of the $J/\psi$ yield at RHIC for two
    centrality  
bins.  The calculations are predictions of the statistical hadronization
model, performed for the nominal pQCD charm production 
cross 
section (continuous line with the band denoting the systematic errors of the 
cross section). The  data are from the PHENIX collaboration
\cite{phe1}. For details see text.} 
  \label{aa_fig4}
\end{figure}

The resulting rapidity dependence of the $J/\psi$ yield for Au-Au collisions
at top RHIC energy is shown in Fig.~\ref{aa_fig4} for two centrality bins.
The PHENIX data \cite{phe1} are well described by the model calculations for
the central value of the pQCD charm cross section.  Since the rapidity
distributions for open charm production are rather wide \cite{cac1}, no visible
narrowing of the calculated J/$\psi$ rapidity distributions is observed, in
contrast to predictions within the kinetic model \cite{the3}. Obviously, the
agreement observed depends sensitively on the magnitude and rapidity
dependence of the open charm cross section and a direct measurement of these
quantities is very important.

Next we focus on the rapidity and centrality dependence of the nuclear
modification factor R$_{AA}$ which has recently been calculated also in the
statistical hadronization model \cite{aa3}. For this purpose, R$_{AA}$ is
defined as 

\be 
R_{AA}^{J/\psi}= \frac{\ud N_{J/\psi}^{AuAu}/\ud
  y}{N_{coll}\cdot\ud N_{J/\psi}^{pp}/\ud y} 
\ee 
and relates the charmonium
yield in nucleus-nucleus collisions to that expected for a superposition of
independent nucleon-nucleon collisions.  Here, $\ud N_{J/\psi}/\ud y$ is the
rapidity density of the $J/\psi$ yield integrated over transverse momentum and
$N_{coll}$ is the number of binary collisions for a given centrality class.
This definition of the modification factor is essentially equivalent to the
J/$\psi$ modification factor employed previously by the NA50 collaboration at
top SPS energy \cite{na50}. 

Important in the evaluation of the nuclear modification factor are also data
on J/$\psi$ production in pp collisions. For experiments at RHIC energy
($\sqrt{s_{NN}}$=200 GeV) we use the recently released data by the PHENIX
experiment \cite{phe2}. For LHC energy we extrapolate the cross section for
J/$\psi$ production in $\bar p p$ collisions measured at the Tevatron
\cite{cdf}.

\begin{figure}[ht]
  \centering\includegraphics[width=.99\textwidth]{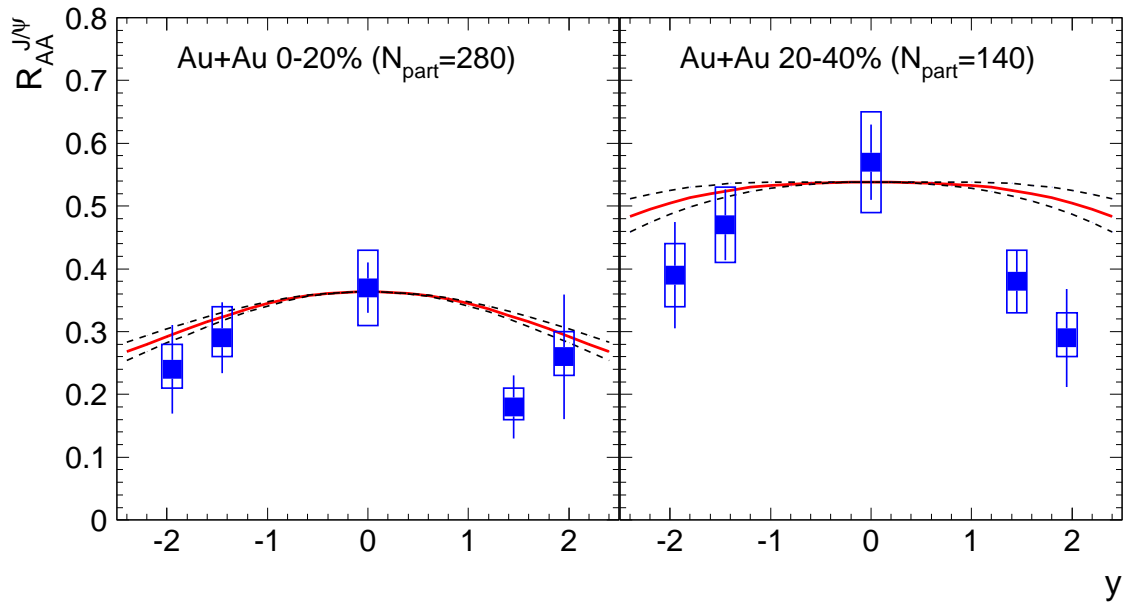}
  \caption{Rapidity dependence of $R_{AA}^{J/\psi}$ for two centrality 
classes. The data (symbols with errors) are compared to calculations (lines).
For the data \cite{phe1}, the error bars show the statistical and 
uncorrelated systematic errors added in quadrature, while the correlated 
systematic errors are represented by the boxes.
Note that a global systematic error of the order of 10\%  has to be 
additionally applied.
The dashed lines denote the error of the gaussian width of the $J/\psi$ 
distribution in pp collisions (see text).
}
\label{aa_fig1}
\end{figure}

In Fig.~\ref{aa_fig1} we present the calculated rapidity dependence of
$R_{AA}^{J/\psi}$ along with the PHENIX experimental results \cite{phe1}. 
For this calculation, we have fitted the pp measurements \cite{phe2} with
a gaussian, with a resulting width in rapidity $\sigma_y=1.63\pm 0.05$
($\chi^2/N_{df}$=4.5/8). As an aside, we note that the fitted gaussian is 
very close to the shape of the rapidity distribution of the pQCD charm cross 
section \cite{cac1}.
Our calculations reproduce quantitatively the $R_{AA}^{J/\psi}$ data,
including the observed larger suppression away from midrapidity. 
We note that this trend is opposite to that expected from the melting model
\cite{satz,satz1}, where $R_{AA}^{J/\psi}$ is constant or exhibits
a minimum at midrapidity. Destruction of charmonia by co-moving hadrons
\cite{gavin_vogt,capella}  should
similarly also lead to the largest suppression at mid-rapidity and, hence,
this mechanism produces results in conflict with the PHENIX data. 

The maximum of $R_{AA}^{J/\psi}$ at midrapidity is in our model due to the
enhanced charmonium production yield at the phase boundary, determined by the
rapidity dependence of the charm production cross section with its maximum at
mid-rapidity. In this sense, the above result constitutes the first
unambiguous evidence for the statistical production of J/$\psi$ at chemical
freeze-out.  In details, our model is in very good agrement with the data for
the central bin (0-20\%), while predicting for the mid-central (20-40\%)
centrality class a somewhat flatter shape than observed in the data.  The
error $\sigma_y$ of the pp data \cite{phe2} used in our model plays a rather
minor role, as denoted by the dashed lines in Fig.~\ref{aa_fig1}.  On the
other hand, the systematic errors of the data including the not exhibited
scale error of the order of 12 \% \cite{phe1} should be taken into account in
a detailed comparison. Since the expected shape in rapidity of the open charm
production cross section at LHC energy is probably even flatter compared to
that at RHIC energy, we expect less variation with rapidity of the nuclear
modification factor for charmonia production as the energy is increased.
On the other hand, $R_{AA}^{J/\psi}$ contains both the pp and AA data, so one
should be open for surprizes. In any case, the rapidity dependence of
$R_{AA}^{J/\psi}$ will be measured in the ALICE experiment \cite{aliceppr} in
the rapidity range $-1 < y < 4$ with precision so that this issue will be
addressed in the near future.   

\begin{figure}[hb]
\vspace{-.3cm}
  \centering\includegraphics[width=.66\textwidth]{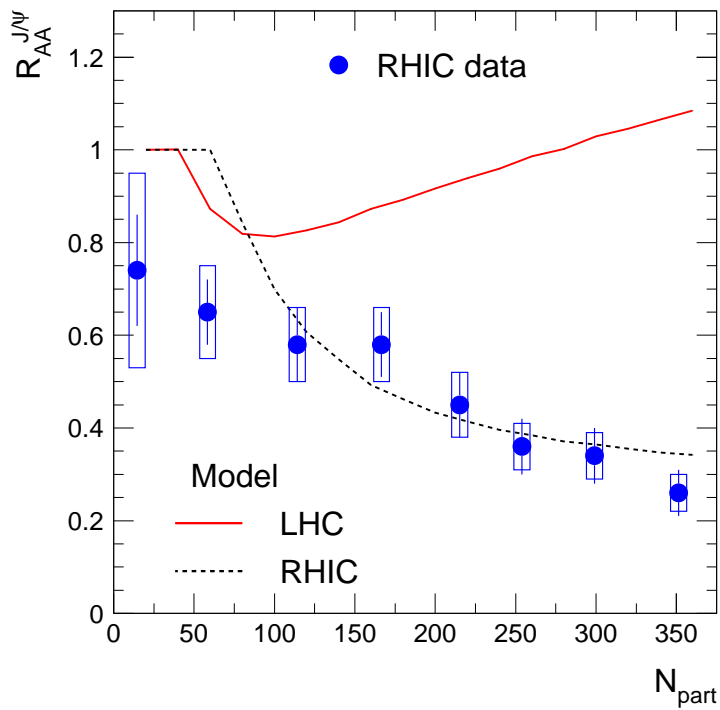}
  \caption{Centrality dependence of the nuclear modification factor 
$R_{AA}^{J/\psi}$ for RHIC and LHC energies. For details see text.} 
\label{aa_fig2}
\end{figure}

The centrality dependence of $R_{AA}^{J/\psi}$ at midrapidity is
shown in Fig.~\ref{aa_fig2}. 
Our calculations approach the value in pp collisions around $N_{part}$=50,
which corresponds to a minimal volume for the creation of QGP 
of 400 fm$^3$ \cite{aa2}.
The model predictions  reproduce very well the decreasing trend versus
centrality  
seen in the RHIC data \cite{phe1}. 
We have not included in our calculations the smearing in $N_{part}$ 
due to finite resolution in the experimental centrality selection.
This effect would lead to a better agreement with data for peripheral
collisions.
Note that in the  statistical hadronization model  the centrality dependence
of the nuclear 
modification factor is a consequence of the still rather moderate 
rapidity density of initially produced charm quark pairs 
($\ud N_{c\bar{c}}/\ud y$=1.6) at top RHIC energy, implying that canonical
thermodynamics has to be used to compute the charm quark fugacity factor in
the charm balance equation \cite{aa2}. 

\begin{figure}[hb]
\vspace{-.3cm}
  \centering\includegraphics[width=.66\textwidth]{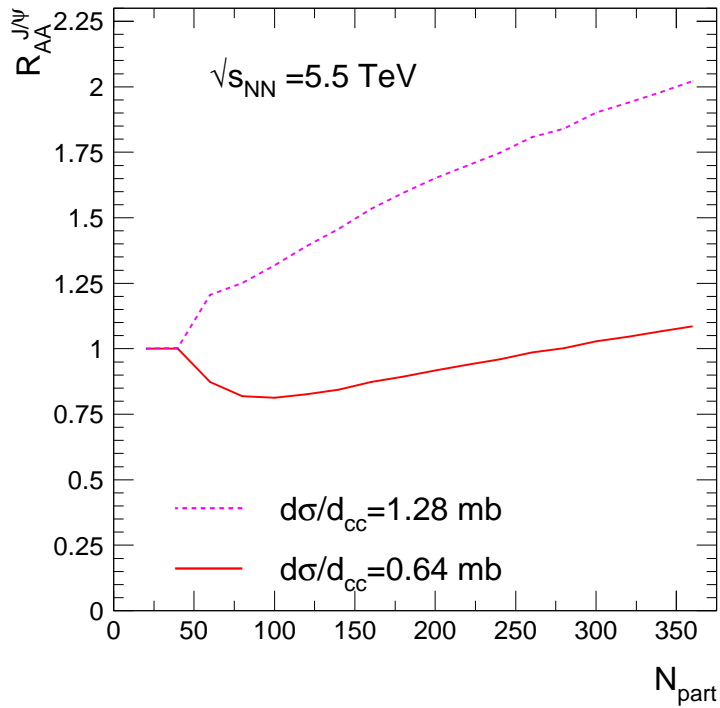}
  \caption{Expected centrality dependence of the nuclear modification factor  
$R_{AA}^{J/\psi}$ for the nominal charm  production cross section and for a
cross section enhanced by a factor of
2 compared with current pQCD calculations.} 
\label{enhance}
\end{figure}

In contradistinction, at the much higher LHC energy, $\sqrt{s_{NN}}$=5.5 TeV, 
the charm production cross section is expected to be about an order
of magnitude larger \cite{rv1,aa2}.
In this case, the canonical correction is sizable only for peripheral
collisions. As a result, a totally opposite trend as a
function of centrality is predicted, see Fig.~\ref{aa_fig2}, with $R_{AA}$
exceeding 1 for central collisions.
A significantly larger enhancement of 2  is obtained if the charm
production cross section is a factor 2 larger than presently assumed, as is
exhibited in Fig.~\ref{enhance}.

In summary, we have presented a brief discussion of the various mechanisms
currently proposed to understand charmonium production in ultra-relativistic
nucleus-nucleus collisions. The main emphasis of this paper is on the
statistical hadronization model. This model has been further developed
recently \cite{aa2,aa3} to include charmonium rapidity and transverse momentum
distributions and a description of the nuclear modification factor
$R_{AA}^{J/\psi}$.  By an analysis of the rapidity dependence of this nuclear
modification factor, for which data were recently published by the PHENIX
collaboration, we have identified, for the first time, a clear signal for
production of charmonia due to statistical hadronization at the phase
boundary. Predictions using this model also describe well the measured
decrease with centrality of $R_{AA}^{J/\psi}$ at RHIC energy. Extrapolation to
LHC energy leads, contrary to the observations at RHIC, to a J/$\psi$ nuclear
modification factor increasing with collision centrality and exceeding 1 for
central collisions. While the exact amount of enhancement will depend on the
precise energy dependence of the open charm production cross section, the
trend is a robust prediction of the model. If the predicted centrality and
rapidity dependence is observed, this would be a striking fingerprint for the
deconfinement of heavy quarks in the QGP. Data from the LHC will be decisive
in settling this important issue and all three large experiments (ALICE,
ATLAS, CMS) are planning to measure charmonium production in the first heavy
ion run of the LHC.


\vspace{0.5cm}
\begin{thebibliography}{99}
\bibitem{satz} T. Matsui, H. Satz, Phys. Lett. B 178 (1986) 416.
\bibitem{satz1} H. Satz, J.Phys. J. Phys. {\bf G32} (2006) R25.
\bibitem{blaizot} J.P. Blaizot, J.Y. Ollitrault, \PRD {39} (1989) 232.
\bibitem{pbm1} P. Braun-Munzinger, J. Stachel,
\Journal{\PLB}{490}{2000}{196} [nucl-th/0007059];
\Journal{\NPA}{690}{2001}{119c} [nucl-th/0012064].
\bibitem{aa1} A. Andronic, P. Braun-Munzinger, K. Redlich, J. Stachel, 
Phys. Lett. B 571 (2003) 36 [nucl-th/0303036].
\bibitem{aa2} A. Andronic, P. Braun-Munzinger, K. Redlich, J. Stachel,
  Nucl. Phys. {\bf A} (submitted), nucl-th/0611023.
\bibitem{wetterich} P. Braun-Munzinger, J. Stachel, C. Wetterich,
  Phys. Lett. {\bf B596} (2004) 61.
\bibitem{rapp} L. Grandchamp and R. Rapp, Nucl. Phys. {\bf A709} (2002) 415.
\bibitem{ceres_prl} D. Adamova et al., CERES collaboration,
  Phys. Rev. Lett. {\bf 90} (2003) 022301.
\bibitem{the1} R.L. Thews, M. Schroedter, J. Rafelski, \Journal{\PRC}{63}
{2001}{054905} [hep-ph/0007323].
\bibitem{the3} R.L. Thews, M.L. Mangano, Phys. Rev. C 73 (2006) 014904
[hep-ph/0505055]. 
\bibitem{yan} L. Yan, P. Zhuang, N. Xu, Phys. Lett. B 607 (2005) 107
[nucl-th/0411093]; Phys. Rev. Lett. 97 (2006) 232301 [nucl-th/0608010].

\bibitem{cac} M. Cacciari, P. Nason, R. Vogt, Phys. Rev. Lett. 95 (2005) 
122001 [hep-ph/0502203].
\bibitem{cac1} M. Cacciari, private communication.
\bibitem{rv1} R. Vogt, Int. J. Mod. Phys. E 12 (2003) 211 [hep-ph/0111271].
\bibitem{phe1} A. Adare et al. (PHENIX collaboration), nucl-ex/0611020.


\bibitem{aa3} A. Andronic, P. Braun-Munzinger, K. Redlich, J. Stachel,
  submitted to Phys. Lett. B (Jan. 2007), nucl-th/0701079.


\bibitem{na50} B. Alessandro et al. (NA50 collaboration), Eur. Phys. J. C 39
  (2005) 335 
[hep-ex/0412036].

\bibitem{phe2} A. Adare et al. (PHENIX collaboration) hep-ex/0611020;

\bibitem{cdf} D. Acosta et al. (CDF collaboration), Phys. Rev. Lett. 91 (2003)
  241804 [nucl-ex/0409028];  G. Pauletta (for the D{\O} and CDF
  collaborations), J. Phys. G 31 (2005) S817.





\bibitem{gavin_vogt} S. Gavin, R. Vogt, Phys. Rev. Lett. {\bf 78} (1997) 1006.

\bibitem{capella} A. Capella, E.G. Ferreiro, Eur. J. Phys. {\bf C42} (2005)
  419. 

\bibitem{aliceppr} B. Alessandro et al., ALICE collaboration, J. Phys. {\bf
    G30} (2004) 1517; F. Carminati et al. ALICE collaboration, J. Phys. {\bf
    G32} (2006) 1295








\end{thebibliography}
\end{document}